\documentclass[11pt,a4paper]{article}

\usepackage[T1]{fontenc}
\usepackage[utf8]{inputenc}
\usepackage{lmodern}
\usepackage{microtype}
\usepackage[margin=1in]{geometry}
\usepackage{amsmath}
\usepackage{booktabs}
\usepackage{array}
\usepackage{enumitem}
\usepackage{listings}
\usepackage{xcolor}
\usepackage[colorlinks=true,linkcolor=blue!45!black,citecolor=blue!45!black,urlcolor=blue!45!black]{hyperref}
\usepackage{url}
\hypersetup{
  pdftitle={Caleta: An Evidence-Oriented Trust Layer for Reproducible Local Large Language Model Inference},
  pdfauthor={Joan-Marc Fisa},
  pdfsubject={Non-peer-reviewed technical preprint on reproducible local LLM inference},
  pdfkeywords={local inference, large language models, reproducibility, privacy engineering, benchmarking, llama.cpp}
}

\setlist{nosep}
\newcommand{\softwareversion}{0.1.0}
\newcommand{\paperdate}{31 October 2026}
\newcommand{\schema}[1]{\texttt{#1}}
\newcommand{\artifact}[2]{\href{https://github.com/BiVelio/caleta/blob/evidence-v1.1/#1}{#2}}

\title{Caleta: An Evidence-Oriented Trust Layer for\\Reproducible Local Large Language Model Inference\\[0.6em]
\large Technical Preprint --- Software Validation Manuscript}
\author{Joan-Marc Fisa\\
BiVelio Inc\\
\texttt{jmfisa@bivelio.com}}
\date{\paperdate}

\begin{document}
\maketitle

\begin{center}
\small
Copyright \copyright\ 2026 BiVelio Inc. This manuscript is licensed under \href{https://creativecommons.org/licenses/by/4.0/}{Creative Commons Attribution 4.0 International (CC BY 4.0)}.\\[0.8em]
Corresponding author: Joan-Marc Fisa (\texttt{jmfisa@bivelio.com}).\\[0.8em]
Software reference implementation: Caleta v\softwareversion, Apache-2.0.\\[0.8em]
Code and evidence: \url{https://github.com/BiVelio/caleta}.\\[0.8em]
Zenodo DOI: \href{https://doi.org/10.5281/zenodo.23839381}{10.5281/zenodo.23839381}.\\[0.8em]
\textbf{Status:} non-peer-reviewed technical preprint.
\end{center}

\begin{abstract}
Local large language model (LLM) inference is often described as private, verified, or reproducible without separating the technical claims behind those terms. This paper presents \textbf{Caleta}, an Apache-2.0 process supervisor and evidence layer around llama.cpp rather than a new inference engine. Caleta combines versioned manifests, exact-size and SHA-256 verification, controlled backend construction, a restricted OpenAI-compatible proxy, prompt-free execution records, benchmark reports, and a small conformance suite. Its contribution is the composition of these mechanisms into explicit contracts that distinguish integrity from authenticated provenance and privacy controls from operating-system isolation. We evaluate the implementation through static checks, \artifact{experiments/software-validation/results/table1-software-validation.json}{22 unit tests}, deterministic end-to-end tests on Ubuntu, macOS, and Windows, and a pinned real-model experiment on Windows CPU using Qwen2.5-0.5B-Instruct Q4\_K\_M and llama.cpp b11981. The real experiment validates artifact verification, startup, model listing, non-streaming and server-sent-event chat, and conformance. Three independent benchmark invocations produced fifteen successful warm requests with no failures. In this single configuration, median time to first token was \artifact{experiments/real-gguf-qwen2.5-0.5b/results/summary.json}{18.066 ms} and median decode throughput was \artifact{experiments/real-gguf-qwen2.5-0.5b/results/summary.json}{54.394 tokens/s}. These descriptive measurements do not establish cross-engine superiority, output quality, energy efficiency, or hardware-general performance. Raw JSON, commands, and environment metadata are published. Remaining challenges include signed manifests, time-of-check/time-of-use replacement, operating-system containment, proxy-overhead measurement, and broader experimental matrices.
\end{abstract}


\clearpage
\section{Introduction}

The deployment of LLMs on local machines is motivated by latency, cost, data-control, availability, and operational-sovereignty requirements. Open-source runtimes such as llama.cpp, MLX-based systems, Ollama, Uzu, and BaseRT address important parts of this problem, including hardware coverage, model packaging, ease of use, and optimized execution \cite{llamacpp,mlx,ollama,uzu,basert}. However, the ability to execute a model locally does not by itself answer several operational questions: Which bytes were executed? Which backend and runtime options were active? Which interfaces were exposed? What was measured, and which runs failed? What evidence remains after execution?

These questions are often conflated under broad claims of privacy or verification. A cryptographic digest can establish correspondence between an artifact and an expected value, but not the identity of the publisher who supplied that value. An offline backend flag can request network inactivity, but it is not equivalent to an operating-system firewall. A local HTTP listener can avoid wide-area exposure while remaining reachable by other local processes. Similarly, a throughput number without artifact identity, generation parameters, cold/warm separation, and failed-run accounting is insufficient to reproduce an experiment.

Caleta addresses this narrower but consequential layer around the inference engine. It deliberately does not implement tensor kernels, a model format, or a new serving API. Instead, it wraps llama.cpp's \texttt{llama-server} with a small trust boundary and emits versioned evidence. This choice preserves GGUF interoperability \cite{gguf} and allows the measurement and conformance contracts to be reused independently of the supervisor.

The manuscript makes three deliberate evidentiary restrictions. First, it reports only tests actually executed on the reference implementation. Second, it separates deterministic mock-based orchestration tests from a real-model experiment on one Windows CPU host. Third, it treats authenticated provenance, OS-level containment, cross-engine comparison, and broader real-model matrices as future work rather than implied properties of the current system.

\section{Research Questions and Contributions}

We organize the work around three research questions:

\begin{description}[leftmargin=1.5em,style=nextline]
\item[RQ1:] How can a local inference launcher make artifact integrity and runtime configuration explicit without replacing the inference engine?
\item[RQ2:] Which operational and benchmarking evidence can be emitted without persisting prompts and responses by default?
\item[RQ3:] How can deterministic tests and a real-model experiment complement each other while keeping their evidentiary scopes distinct?
\end{description}

The paper offers four contributions:

\begin{enumerate}
\item A composable architecture that combines artifact verification, backend supervision, restricted proxying, execution records, benchmarking, and conformance under explicit security boundaries.
\item Versioned manifest and report contracts that separate declared provenance, observed integrity, runtime configuration, and measurement limitations.
\item Benchmark definitions that preserve raw failures, distinguish cold startup from warm requests, and avoid converting unavailable measurements into zero-valued results.
\item A two-layer evaluation comprising cross-platform deterministic tests and a pinned real-GGUF experiment with raw reports, commands, environment metadata, and a threat-to-validity analysis.
\end{enumerate}

The contribution is therefore not a universal performance improvement or a formal proof of privacy. It is a reference design for bounded, inspectable claims in local inference systems.

\section{Background and Related Work}

\subsection{Local inference runtimes}

llama.cpp provides broad GGUF and hardware support and includes an HTTP server with offline, authentication, and cross-origin resource sharing (CORS) options \cite{llamacpp}. Caleta does not assume that these upstream controls are absent. Its purpose is to construct a consistent subset of configuration, remove selected inherited overrides, restrict externally reachable routes, and produce execution and measurement evidence around that configuration.

Ollama emphasizes installation and model-distribution ergonomics \cite{ollama}; MLX LM integrates local inference with Apple Silicon and the Hugging Face ecosystem \cite{mlx}; Uzu implements a Rust inference engine with platform-specific optimization and a broader application ecosystem \cite{uzu}. BaseRT demonstrates the importance of native Metal optimization and phase-specific performance evaluation on Apple Silicon \cite{basert}. These projects primarily address execution, optimization, packaging, or developer experience. Caleta is complementary: its schemas and conformance checks could be implemented by another runtime without adopting its process supervisor.

The BaseRT preprint also illustrates why performance claims must be scoped by hardware, model family, quantization, and inference phase. Its results concern the configurations evaluated by its authors. They are neither replicated in this study nor used as evidence for Caleta's performance.

\subsection{Integrity, provenance, and attestation}

Secure software distribution requires more than file hashing. SHA-256 provides a deterministic digest used here to compare artifact contents against a manifest \cite{fips180-4}. The security of that comparison depends on how the expected digest was obtained. Supply-chain frameworks such as SLSA distinguish provenance and build properties from the simplified notion that a checksum alone establishes trust \cite{slsa}.

Caleta v\softwareversion\ implements integrity comparison and provenance declaration. It does not implement signed manifests, transparency logs, remote attestation, or independently signed execution records. We reserve \emph{authenticated provenance} for mechanisms that establish the identity of the party supplying an artifact or manifest through a trusted channel.

\subsection{Reproducibility and transparency in AI}

Reproducibility research distinguishes the artifacts, computational environment, methodology, and analysis needed to obtain similar results \cite{gundersen,pineau}. Model cards and datasheets provide structured documentation for trained models and datasets, including their intended uses and limitations \cite{modelcards,datasheets}. Caleta operates at a complementary layer: its manifest and execution reports describe what was run and how, not how the model was trained, its full evaluation profile, or its suitability for a task. A Caleta manifest therefore should accompany, rather than replace, model and dataset documentation.

\subsection{Reproducible performance evaluation}

Local inference comprises at least model initialization, prompt processing, and autoregressive decoding. Combining these stages into one number obscures different system behaviors. Meaningful comparisons must hold artifact identity, quantization semantics, prompts, templates, decoding parameters, backend revisions, hardware, and power conditions as constant as practical \cite{pineau}. They must also distinguish cold starts from warm requests and describe how tokens and timing events are counted.

Caleta operationalizes part of this methodology in JSON. It does not measure output quality, energy, thermal throttling, GPU memory, or pure kernel time. These omissions are represented as limitations rather than estimated values.

\section{Design Principles}

The architecture follows five principles:

\paragraph{Composition over replacement.} The reference backend remains llama.cpp. Caleta does not duplicate its kernels, model conversion, or hardware backends.

\paragraph{Verification before execution.} \texttt{serve} and \texttt{bench} refuse to start after a missing artifact, size mismatch, digest mismatch, or read failure.

\paragraph{Least exposed interface by default.} Both the public proxy and backend default to loopback. The backend is never intentionally exposed on a public listener. Only health, model listing, and chat completions are proxied.

\paragraph{Evidence with bounded meaning.} Every contract names what it records and what it does not establish. A verification report is labeled \texttt{integrity\_only}; execution records are unsigned operational evidence, not attestations.

\paragraph{Reproducibility without prompt retention by default.} Configuration and measurements are retained, while request and response bodies are not logged. Benchmarks store a prompt digest and length unless the operator explicitly requests prompt inclusion.

\section{System Architecture}

\subsection{Components and data flow}

Figure~\ref{fig:architecture} shows the principal boundary. A client communicates with the Caleta proxy. The proxy replaces external authorization with a random per-launch internal credential and forwards only supported requests to llama-server. The backend binds to IPv4 loopback and receives a verified local GGUF path.

\begin{figure}[ht]
\centering
\begin{lstlisting}
OpenAI-compatible client
          |
          | public bearer credential when configured
          v
    Caleta HTTP proxy ---- execution evidence / CORS policy
          |
          | random internal credential; loopback only
          v
      llama-server ------- --offline; no Web UI; controlled arguments
          |
          v
 verified local GGUF artifact
\end{lstlisting}
\caption{Caleta process and evidence boundary. The diagram does not imply OS-level sandboxing.}
\label{fig:architecture}
\end{figure}

The implementation separates manifest parsing, verification, backend lifecycle, HTTP serving, execution records, diagnostics, benchmarking, and conformance. This decomposition permits unit testing without a model and keeps the command-line interface from becoming the sole integration point.

\subsection{Manifest and artifact verification}

A manifest declares a schema identifier, model identifier, license, source, artifact path, expected digest and size, backend kind, and runtime parameters. Relative artifact and explicit backend paths resolve against the manifest directory rather than the caller's working directory. The source and license are declarations; they are not inferred from file contents.

For artifact bytes $a$, declared digest $h$, and declared byte length $b$, execution proceeds only if
\begin{equation}
|a| = b \quad \land \quad \operatorname{SHA256}(a) = h.
\end{equation}
Verification streams data in 1~MiB blocks, so artifact size does not determine verifier memory usage. When a backend digest is pinned, an explicit backend path is required to avoid treating ambiguous \texttt{PATH} resolution as precise provenance. Optional `backend.dependencies` entries pin individual native dependencies by relative path and digest; an empty list does not claim a fully verified native distribution.

This check does not establish that the manifest is truthful, that the model is safe to parse, or that the file cannot be replaced between verification and backend loading. Signed manifests and immutable or otherwise protected storage are therefore separate requirements.

\subsection{Backend supervision}

Before launch, Caleta removes inherited \texttt{LLAMA\_ARG\_*} and \texttt{LLAMA\_API\_KEY} variables that could alter the declared boundary. It generates a fresh 32-byte random internal credential, passes it through the backend environment, and constructs arguments for model, loopback host, ephemeral port, alias, offline mode, disabled Web UI, context size, GPU layers, and optional thread count.

The manifest validator rejects additional arguments that would replace the model or listener, alter credentials, or enable agent tools. This is a reviewed policy for known options, not a universal proof against every future upstream flag. Readiness requires an authenticated health response within a timeout. Backend \texttt{--version} has a 10-second cap. Benchmark and conformance clients use 120-second timeouts; readiness health polls use shorter per-attempt deadlines within the configured startup timeout. Premature child exit and readiness timeout are errors. Shutdown handles SIGINT and SIGTERM, terminates and waits for the child, checks its final state, and uses kill-on-drop as an additional process-lifetime safeguard. The `/health` endpoint reports actual backend liveness and returns 503 after child exit.

Two residual races remain: another local process may bind a reserved ephemeral port before the child uses it, and an artifact may be replaced after verification. Hostile-local-user deployments require stronger OS-specific mechanisms.

\subsection{Restricted HTTP proxy}

The public server exposes only \texttt{/health}, \texttt{/v1/models}, and \texttt{/v1/chat/completions}, a deliberately small subset of the OpenAI-compatible API \cite{openai}. Model and chat endpoints require a public bearer key of at least sixteen characters by default on every listener, including loopback; unauthenticated loopback requires an explicit development opt-in. The health endpoint does not expose prompts or generation. Credential comparison is constant-time. The proxy accepts at most 32 concurrent requests and times out upstream calls after 120 seconds. Chat requires \texttt{application/json}; model and chat requests validate loopback hosts and browser origins. Request bodies are limited to 16~MiB; responses, including server-sent events (SSE), are streamed rather than collected. Error response bodies are omitted from reports and logs.

The proxy removes hop-by-hop headers and all upstream CORS response headers. Only Caleta's explicit origin allowlist can produce CORS headers. Wildcard origins are rejected. The server provides plain HTTP; remote deployments require a TLS reverse proxy, SSH tunnel, or an appropriately controlled private network. CORS is not an authentication mechanism and does not protect non-browser clients.

\section{Operational Evidence and Measurement Contracts}

\subsection{Execution records}

Optional execution records use the schema \schema{caleta.dev/execution/v1}. They capture version, model identity, paths, artifact and backend digests, listener configuration, loopback/offline flags, CORS policy, arguments, and limitations. Prompt and response bodies are excluded. The record is generated by the same process whose behavior it describes and is unsigned; it supports inspection and reproduction but does not prove execution to a third party.

\subsection{Benchmark methodology}

A benchmark verifies the artifact, starts one backend, measures cold startup to authenticated readiness, executes warm-up requests, and then measures warm requests in that process. Failed measured requests remain in the report and are excluded only from aggregates of successful measurements. Any measured failure also produces a nonzero exit status after the report is written. Warm-up failures are logged but are not retained as raw per-run records. A report is written even when all measured requests fail.

For a request sent at $t_0$, the first content-bearing SSE event arrives at $t_1$, and the last such event arrives at $t_n$. Time to first token is defined operationally as
\begin{equation}
\operatorname{TTFT} = t_1-t_0.
\end{equation}
It includes local HTTP and queueing overhead and ignores empty role-only events. If the backend reports $c$ completion tokens and $c\geq2$, decode throughput is
\begin{equation}
\operatorname{decode\_tps}=\frac{c-1}{t_n-t_1}.
\end{equation}
Observed prefill throughput is reported prompt tokens divided by TTFT; it is not a pure kernel measurement. Repeating an identical prompt within one backend process may also reuse cached prompt state, so this value must not be interpreted as cold prefill performance. Inter-event latency is computed over arrivals of content-bearing SSE events. Because backend chunking need not map one event to one tokenizer token, this metric is not claimed to be exact per-token latency in all backends.

The schema \schema{caleta.dev/benchmark/v1} stores raw runs, token usage, estimated-token flags, cold startup, configuration, environment, failures, and limitations. On Linux, main-process resident memory is sampled approximately every 50~ms through \texttt{/proc/PID/status}; this excludes GPU memory and child processes and can miss short peaks. Missing metrics remain null.

\subsection{Conformance}

The conformance command checks a deliberately small OpenAI-compatible contract: model listing returns a data array, non-streaming chat contains message content in the expected structure, and streaming chat uses the base MIME \texttt{text/event-stream}, contains structural JSON delta events, and terminates with an exact \texttt{data: [DONE]} event. The schema is \schema{caleta.dev/conformance/v1}. Passing does not certify tool calls, structured output, multimodality, cancellation, backpressure, fragmented SSE handling, or complete OpenAI compatibility.

\section{Implementation and Reproducibility}

The reference implementation is written in Rust and uses Tokio, Axum, Reqwest, Serde, and SHA-2. The package declares Rust 1.88 as its minimum version and Apache-2.0 as its software license. Model artifacts and llama.cpp retain their respective licenses. Caleta does not bundle a model or llama.cpp binary.

A minimal local workflow is:

\begin{lstlisting}
cargo build --release --locked
caleta doctor --backend /path/to/llama-server
caleta verify model.toml
caleta serve model.toml --record-out execution.json
caleta bench model.toml --output benchmark.json \
  --runs 5 --warmup 1 --max-tokens 64 --seed 42
caleta conformance --url http://127.0.0.1:8080 \
  --output conformance.json
\end{lstlisting}

The canonical public artifact location is:
\begin{center}
\url{https://github.com/BiVelio/caleta}
\end{center}
The repository contains the source, lockfile, manifests, and deterministic mock. Future experimental publications must also include raw reports and exact command transcripts.

\section{Evaluation}

\subsection{Experimental scope}

The evaluation has two layers. The first uses a deterministic Python backend that accepts the subset of llama-server arguments used by Caleta, requires the internal credential, and returns fixed protocol responses. Its 45-byte artifact is not a valid GGUF model. This layer tests orchestration and protocol behavior without claiming inference. Local validation used Windows with rustc 1.94.1, Cargo 1.94.1, and Python 3.14.0; hosted CI subsequently covered Ubuntu, macOS, and Windows.

The second layer executes a real Qwen2.5-0.5B-Instruct Q4\_K\_M GGUF through pinned llama.cpp binaries on one Windows CPU host. It tests artifact verification, real model startup, API behavior, conformance, and descriptive inference timing. It does not evaluate output quality, safety, energy, or cross-engine performance.

\subsection{Procedure}

The validation sequence was:

\begin{lstlisting}
cargo fmt --all -- --check
cargo clippy --locked --all-targets -- -D warnings
cargo test --locked
cargo build --locked
CALETA_BIN=target/debug/caleta python3 tests/e2e_mock.py
\end{lstlisting}

The end-to-end scenario constructs a temporary artifact and manifest, computes their digests, and executes verification. It starts the supervised service, checks unauthenticated rejection, performs authorized model listing and streaming chat, and runs the conformance command. It then executes one warm-up and two measured benchmark requests and inspects the resulting JSON. The Unix branch of the scenario validates graceful interruption and the resulting prompt-free execution record; that branch was not executed in the local Windows run. The Windows scenario uses forced process-tree shutdown for the mock and does not claim graceful-shutdown or execution-record coverage.

\subsection{Results}

All reported checks completed successfully after correcting platform-specific implementation and test issues: an unused non-Linux process identifier in the benchmark module rejected by strict Clippy, and Windows handling of TOML paths, executable mock launching, and mock-process shutdown in the test harness. Table~\ref{tab:results} maps outcomes to the claims they support.

\begin{table}[ht]
\centering
\small
\begin{tabular}{>{\raggedright\arraybackslash}p{0.27\textwidth}>{\raggedright\arraybackslash}p{0.18\textwidth}>{\raggedright\arraybackslash}p{0.43\textwidth}}
\toprule
Check & Observed result & Supported interpretation \\
\midrule
Formatting & Pass & Source conforms to Rust formatting. \\
Clippy, all targets & Pass with warnings denied & No emitted lint warnings under the configured toolchain. \\
Unit tests & \artifact{experiments/software-validation/results/windows-latest/unit_tests.json}{22 passed, 0 failed} & Tested manifest, digest, SSE, percentile, authentication, proxy, health, and exposure cases behave as expected. \\
Locked build & Pass & Package builds with the committed dependency resolution. \\
Verification E2E & Pass & Exact size/digest and pinned mock backend are accepted by the workflow. \\
Serve/authentication E2E & Pass & Supervised proxy starts; missing public credentials are rejected; authorized models and chat requests succeed. \\
SSE/conformance E2E & Pass & The tested stream and three conformance contracts pass against the deterministic backend. \\
Benchmark E2E & 2 successful runs & Raw successful-run records and model digest are emitted; no prompt text is stored by default. \\
Mock-based inference claims & Not made & The fixture validates orchestration only; real-model evidence is reported separately below. \\
\bottomrule
\end{tabular}
\caption{\artifact{experiments/software-validation/results/table1-software-validation.json}{Local software-validation outcomes and their bounded interpretation.}}
\label{tab:results}
\end{table}

\subsection{Hosted multi-platform confirmation}

After the local validation, GitHub Actions run \href{https://github.com/BiVelio/caleta/actions/runs/41216696756}{41216696756} completed successfully at source revision \texttt{730a0727b969d7d6b9311d453253d88a0645b9bb}. The Ubuntu, macOS, and Windows test jobs passed, as did the separate formatting, Clippy, and osaudit jobs. The Ubuntu and macOS end-to-end runs included graceful-shutdown execution records; the Windows run used forced mock-process shutdown. The \artifact{experiments/software-validation/results/table1-software-validation.json}{committed JSON and logs} preserve all three runner results, including the 45-byte mock artifact and the 22 passing unit tests. The evidence tag preserves these outputs together with the final manuscript and documentation. This confirms the same mock-based contracts across the three hosted operating systems, but it does not add real-model evidence.

No tokens-per-second result from the mock is scientifically meaningful. Mock timing is intentionally excluded from the performance claims.

\subsection{Real-model procedure}

The real experiment used Qwen2.5-0.5B-Instruct Q4\_K\_M under Apache-2.0, pinned at Hugging Face revision \texttt{9217f5db79a2} \cite{qwenmodel}. The artifact has \artifact{experiments/real-gguf-qwen2.5-0.5b/results/environment.json}{491,400,032 bytes} and this publisher-declared SHA-256:
\begin{lstlisting}
74a4da8c9fdbcd15bd1f6d01d621410d31c6fc00986f5eb687824e7b93d7a9db
\end{lstlisting}
The backend was the Windows CPU asset from llama.cpp release \texttt{b11981}. Both the release archive and executable were hash-pinned. Caleta was built from revision \texttt{539ed3e32b09}. Full identifiers and digests are preserved in the public experiment artifacts.

The host was an MSI MS-7A78 running Windows 10 Home 10.0.19045 with an Intel i7-7700K, four physical cores, and 15.96 GiB RAM. Execution was CPU-only with four threads, a 2,048-token context, \texttt{--jinja}, seed 42, temperature zero, and a 64-token output limit. Power state, thermal state, and unrelated system load were not measured or controlled.

After successful \texttt{doctor} and \texttt{verify} checks, Caleta started the real backend on loopback. The experiment retained health and model-list responses, a non-streaming chat response, a streaming SSE session with usage and \texttt{[DONE]}, and the conformance report. The service and backend were stopped before benchmarking. Three independent benchmark invocations each performed one cold start, two warm-ups, and five measured requests with the same prompt.

\subsection{Real-model results}

Verification passed for the real model and backend. Model listing, non-streaming chat, and SSE streaming completed successfully, and all three conformance checks passed. All fifteen measured benchmark requests succeeded; no failures required exclusion. Table~\ref{tab:realmetrics} reports descriptive aggregates computed from the three raw reports.

\begin{table}[ht]
\centering
\small
\begin{tabular}{>{\raggedright\arraybackslash}p{0.28\textwidth}>{\raggedright\arraybackslash}p{0.25\textwidth}>{\raggedright\arraybackslash}p{0.35\textwidth}}
\toprule
Metric & Observed value & Scope \\
\midrule
Independent cold starts & \artifact{experiments/real-gguf-qwen2.5-0.5b/results/summary.json}{3} & One per benchmark invocation \\
Measured warm requests & \artifact{experiments/real-gguf-qwen2.5-0.5b/results/summary.json}{15} & Five per invocation; zero failures \\
Cold start, median & \artifact{experiments/real-gguf-qwen2.5-0.5b/results/summary.json}{835.371 ms} & Range \artifact{experiments/real-gguf-qwen2.5-0.5b/results/summary.json}{660.720--977.129 ms} \\
TTFT, median & \artifact{experiments/real-gguf-qwen2.5-0.5b/results/summary.json}{18.066 ms} & Range \artifact{experiments/real-gguf-qwen2.5-0.5b/results/summary.json}{14.574--28.004 ms} \\
Decode throughput, median & \artifact{experiments/real-gguf-qwen2.5-0.5b/results/summary.json}{54.394 tokens/s} & Range \artifact{experiments/real-gguf-qwen2.5-0.5b/results/summary.json}{49.028--68.138 tokens/s} \\
Decode sample standard deviation & \artifact{experiments/real-gguf-qwen2.5-0.5b/results/summary.json}{6.617 tokens/s} & Descriptive; not a confidence interval \\
\bottomrule
\end{tabular}
\caption{\artifact{experiments/real-gguf-qwen2.5-0.5b/scripts/aggregate_results.py}{Single-host real-model measurements. They are not cross-engine or hardware-general claims.}}
\label{tab:realmetrics}
\end{table}

Repeated prompts within each backend process may benefit from caching. TTFT includes HTTP and queueing, and observed prefill values are not cold kernel measurements. The experiment does not estimate public-proxy overhead because Caleta's benchmark talks directly to its supervised backend. It also does not measure energy, memory, output quality, safety, or thermal behavior. Retained chat responses provide protocol evidence, not an exact-instruction-following evaluation.

The manifest, commands, environment record, raw JSON reports, SSE session, and aggregate summary are publicly available in the \artifact{experiments/real-gguf-qwen2.5-0.5b/README.md}{real-GGUF experiment directory}. No model or backend binary is redistributed there.

For RQ1, the real experiment confirms that the manifest, verification, supervision, and proxy operate with a pinned GGUF and llama.cpp build, complementing the deterministic tests. For RQ2, the experiment provides raw operational and benchmark evidence without private prompts; hosted Unix runs separately validate prompt-free execution records. For RQ3, the two layers demonstrate why protocol determinism and real-model validation are complementary rather than interchangeable evidence.

\subsection{Artifact traceability}

Every empirical claim in this evaluation is backed by a committed data file or a committed script. Table~\ref{tab:results} is aggregated by \artifact{tests/aggregate_validation.py}{tests/aggregate\_validation.py} from preserved runner outputs, while Table~\ref{tab:realmetrics} is regenerated by \artifact{experiments/real-gguf-qwen2.5-0.5b/scripts/aggregate_results.py}{aggregate\_results.py --check} from the three raw benchmark reports. The \artifact{osaudit.toml}{osaudit configuration} links the paper to these results and instruments, and returns no findings at the evidence tag. This audit establishes mechanical traceability, not independent scientific correctness, external replication, or peer review.

\section{Security and Privacy Analysis}

Table~\ref{tab:threats} summarizes the implemented controls and residual risks. The central analytical requirement is to avoid converting a control in one layer into a guarantee about the whole system.

\begin{table}[ht]
\centering
\small
\begin{tabular}{>{\raggedright\arraybackslash}p{0.23\textwidth}>{\raggedright\arraybackslash}p{0.32\textwidth}>{\raggedright\arraybackslash}p{0.33\textwidth}}
\toprule
Threat or ambiguity & Current control & Residual risk or non-claim \\
\midrule
Artifact substitution & Exact size and SHA-256 before serve/bench & Trusted digest distribution and post-check replacement remain unresolved. \\
Backend substitution & Optional explicit-path digest & Unpinned PATH resolution remains possible and is reported as such. \\
Inherited reconfiguration & Remove selected llama.cpp environment options & Other host variables, drivers, and future upstream options remain relevant. \\
Unauthorized network API use & Bearer authentication on all listeners; explicit local development opt-out & No integrated TLS; secret storage and transport are operator responsibilities. \\
Permissive upstream CORS & Strip upstream CORS; explicit local allowlist & CORS does not authenticate non-browser clients. \\
Backend network access & Loopback listener and offline request & No OS-enforced outbound-network denial. \\
Prompt retention by Caleta & No request/response logs; prompt digest by default & Clients, backend, OS swap, crash reports, and other processes are separate. \\
Misleading execution evidence & Versioned prompt-free JSON & Records are unsigned and are not remote attestations. \\
\bottomrule
\end{tabular}
\caption{Security controls and residual risks in Caleta v\softwareversion.}
\label{tab:threats}
\end{table}

A fully malicious host operating system, privileged local attacker, compromised Caleta binary, or malicious native backend is outside the containment claimed by this version. Model parsing occurs inside llama.cpp, so a matching digest cannot make an untrusted GGUF safe to parse. Loopback also is not a complete security boundary against other local processes.

Accordingly, we define the supported privacy claim narrowly: Caleta adds no telemetry or prompt history, avoids logging request and response bodies, requests offline backend behavior, and limits backend network listening to loopback. It does not claim that the complete host cannot disclose data.

\section{Threats to Validity}

\paragraph{Construct validity.} Integrity, declared provenance, authenticated provenance, privacy, and safety are distinct constructs. The implementation and reports use separate terms, but users could still overinterpret the word ``verified.'' The \texttt{integrity\_only} label and explicit limitations mitigate rather than eliminate this risk.

\paragraph{Internal validity.} The mock backend does not reproduce GGUF parsing, model initialization, tokenization, or hardware behavior. The real experiment covers those mechanisms for one pinned artifact and backend revision, but cannot expose defects specific to other models, versions, or configurations. Repeated prompts may reuse cached state, and the Windows mock uses forced process-tree shutdown, so graceful-shutdown evidence remains platform-specific.

\paragraph{External validity.} Hosted mock-based CI on three operating systems and one real-model Windows CPU experiment cannot establish correctness or performance across backend revisions, model families, hardware accelerators, or adversarial conditions. No macOS or Linux real-model run is claimed. The results should be generalized only through additional pinned experiments.

\paragraph{Reliability.} The committed lockfile, exact source and backend revisions, immutable model revision, publisher digests, commands, environment record, raw JSON, and versioned schemas support repetition of the reported experiment. Absolute local paths in diagnostic evidence and uncontrolled host load can affect exact repetition; timing values are not guaranteed to reproduce.

\paragraph{Performance validity.} Three cold starts and fifteen warm requests provide descriptive evidence for one host, not a stable performance distribution or comparative ranking. The study cannot estimate public-proxy overhead, energy, thermal effects, output quality, or cross-hardware performance. The sample standard deviation is reported without treating it as a confidence interval, and no claims are made about prefill kernels, memory, or quality.

\section{Research Agenda}

The next stage must add evidence rather than broaden marketing language. We identify four ordered directions.

\subsection{Real-model experimental matrix}

A preregistered matrix should pin at least one redistributable GGUF artifact and llama.cpp revision per claimed operating system. Experiments should record model source and digest, quantization, context, prompt template, seed, temperature, hardware, power state, backend options, and Caleta revision. Direct-backend and proxied measurements should estimate overhead with repeated independent runs and uncertainty intervals. Raw successful and failed runs should be published.

\subsection{Authenticated provenance and release evidence}

Signed manifests, protected update channels, release attestations, and independently reproducible builds should be evaluated as separate mechanisms. Acceptance tests must demonstrate signature verification and failure under modification. A software bill of materials may complement these controls but cannot replace them.

\subsection{Operating-system containment}

Network and filesystem restrictions require platform-specific enforcement and adversarial tests. A backend flag must not be reported as a sandbox. The acceptance criterion is a demonstrated blocked outbound connection and denied unauthorized resource access under the stated threat model.

\subsection{Protocol and measurement robustness}

Conformance should expand to fragmented SSE, cancellation, concurrent requests, backpressure, structured outputs, and tool calls only when each contract is implemented and tested. Measurement research should add calibrated energy, GPU and process-tree memory, thermal state, and quantization-quality evaluation without conflating their definitions across platforms.

\section{Conclusion}

Caleta explores a complementary layer in the local-LLM stack: making artifact identity, runtime configuration, exposed interfaces, and measurement methodology explicit and machine-readable. Its design avoids duplicating optimized runtimes and instead composes integrity checks, supervision, restricted proxying, prompt-free operational evidence, benchmarking, and conformance around llama.cpp.

Hosted deterministic tests on three operating systems and a pinned real-GGUF experiment on Windows CPU validate the tested behavior, including real startup, chat, streaming, conformance, and benchmarking. The real measurements are deliberately scoped to one configuration and are not a comparative performance claim. This distinction is central to the contribution: trustworthy local inference requires claims whose scope is visible in both documentation and artifacts. Caleta provides an open implementation and reproducible evidence for that principle, with future work needed on broader matrices, authenticated provenance, containment, and measurement.

\section*{Author and Intellectual Property}

\textbf{Author and corresponding author:} Joan-Marc Fisa, BiVelio Inc, \texttt{jmfisa@bivelio.com}.\\
\textbf{Manuscript copyright and license:} Copyright \copyright\ 2026 BiVelio Inc. Distributed under \href{https://creativecommons.org/licenses/by/4.0/}{CC BY 4.0}; attribution must identify Joan-Marc Fisa as author and BiVelio Inc as copyright holder.\\
\textbf{Reference software license:} Apache License 2.0. Third-party software and model artifacts retain their own licenses. The software license does not grant trademark rights in Caleta or BiVelio.

\section*{Data and Code Availability}

The reference implementation, test harness, documentation, and release runbook are available at \url{https://github.com/BiVelio/caleta}. The real-model manifest, commands, environment record, raw reports, and aggregate summary are available in the \artifact{experiments/real-gguf-qwen2.5-0.5b/README.md}{experiment artifacts directory}. This manuscript version is assigned DOI \href{https://doi.org/10.5281/zenodo.23839381}{10.5281/zenodo.23839381}; earlier versions remain available through \href{https://doi.org/10.5281/zenodo.23761754}{10.5281/zenodo.23761754}.

\raggedright

\end{document}